\documentclass[12pt,reqno]{amsart} 
\usepackage{amsmath,amssymb,amsthm,amsfonts}
\usepackage{natbib}
\usepackage{graphicx}
\usepackage{relsize}
\usepackage{catchfilebetweentags}
\usepackage{layout}
\usepackage{bm}
\usepackage{xcolor}
\usepackage{subcaption} 
\usepackage[shortlabels]{enumitem}
\usepackage{caption}
\usepackage{float}
\usepackage{multirow}
\usepackage{multicol}
\usepackage[margin=1in]{geometry} 
\usepackage[onehalfspacing,nodisplayskipstretch]{setspace} 
\usepackage{mathtools}
\usepackage{threeparttable}

\newtheorem{theorem}{Theorem}
\newtheorem{lemma}{Lemma}
\newtheorem{definition}{Definition}
\newtheorem{proposition}{Proposition}
\newtheorem*{example*}{Example}

\newtheorem{assumption}{Assumption}

\newcommand{\m}[1]{\mathcal{#1}}

\newcommand{\mme}{\mathbb{E}}
\newcommand{\mmi}{\mathbb{I}}
\newcommand{\mmp}{\mathbb{P}}

\newcommand{\mJ}{\m{J}}

\makeatletter
\def\old@comma{,}
\catcode`\,=13
\def,{%
  \ifmmode%
    \old@comma\discretionary{}{}{}%
  \else%
    \old@comma%
  \fi%
}
\makeatother

\begin{document}
\pagestyle{plain}

\title[Roy+RD]{Roy Models in Centralized Assignment Mechanisms}
\author{\tiny Marinho Bertanha, Marc Henry, Margaux Luflade, Isma\"{e}l Mourifi\'e, Goya Razavi}
\address{University of Notre Dame, Penn State, The University of Pennsylvania, Washington University in St. Louis, Norwegian School of Economics}
\thanks{
This version: November 26, 2026. 
Corresponding author: Marinho Bertanha: Department of Economics, University of Notre Dame, 3060 Jenkins and Nanovic Halls, Notre Dame, IN 46556, USA.
Generative AI tools were utilized during the development of this paper for prose editing, conceptual brainstorming, literature search, and code drafting/debugging.
All outputs were rigorously verified by the authors, who bear ultimate responsibility for the manuscript's text, theoretical proofs, and conclusions.
}

\maketitle

\begin{center}
\textbf{Abstract}
\end{center}

We propose an empirical framework that combines local causal evidence with economic restrictions on program choice to learn about global returns to university programs and the effects of centralized admissions reforms. Admission cutoffs provide identifying information about students at admission margins, whereas policy reforms can change assignments for students away from those margins. Extrapolating local effects to these students need not produce predictions consistent with their observed choices under an economic model of selection. Our approach uses a Roy model to extract restrictions on unobserved potential outcomes from students’ submitted rankings and combines these restrictions with admission cutoffs, realized labor market outcomes, and monotonicity in admission scores. We derive bounds on global causal and counterfactual parameters that incorporate both the local identifying information and the behavioral information contained in choices.  The framework supports evaluation of policies such as affirmative action and capacity reallocations while making explicit the assumptions needed to connect local evidence to global policy conclusions.

\vskip15pt


\vskip12pt


\normalsize

\section{Introduction}

A growing literature exploits centralized assignment mechanisms to identify the causal effects of educational  choices on later labor market outcomes. A key insight is that stable assignments admit a cutoff characterization: each program has an admission threshold, and students are assigned to their preferred program among those for which they qualify. This characterization, developed by \cite{azevedo2016}, makes admission scores and cutoffs informative about students' feasible choices and provides a foundation for regression discontinuity designs. Under appropriate continuity conditions, changes in eligibility at admission cutoffs identify causal effects for students at the admission margin. These designs have generated evidence on the returns to institutions and fields of study while requiring limited restrictions on the economic behavior underlying students' choices. Applications using admission thresholds include \cite{kirabo_ej2010}, \cite{hoekstra2009}, \cite{hastingsneilsonzimmerman2013}, \cite{zimmer2014}, \cite{kirkeboen2016}, \cite{otero2021}, \cite{abdul_ecta2022}, \cite{razavi2026major} and \cite{blm_arxiv}.

Many policy questions, however, require information beyond these local comparisons. Changes in program capacities, admissions priorities, or assignment rules can reallocate students across several programs. Because assignments are interconnected, a reform can affect students beyond its intended beneficiaries, and a nonmarginal reform can change assignments for students whose scores are away from existing cutoffs. Evaluating its aggregate and distributional consequences therefore requires information about potential outcomes throughout the population whose assignments change. Similarly, returns to programs of study  depend on treatment effects beyond those identified at admission margins. Local causal effects provide valuable information for these questions, but generally do not determine the global parameters of interest.

This limitation is addressed by an emerging literature that includes \cite{bertanha_joe2020}, \cite{cattaneo2020random}, and \cite{imai_2025}.  They develop methods to go beyond regression discontinuity cutoffs by using observed choices of students away from the cutoffs, which are available in centralized assignment mechanisms. In this vein, we propose an approach that combines local identifying information with economic restrictions on selection to bound global causal and counterfactual parameters. The economic restrictions translate the observed rank-order lists into restrictions on unobserved potential labor market outcomes. We combine these restrictions with information from the assignment mechanism, realized outcomes, and continuity at the cutoffs.

We develop this approach in a Roy model of college and labor market sorting. The Roy model provides a natural framework for incorporating self-selection based on comparative advantage: students' preferences over some programs reflect the labor market outcomes they anticipate in those programs. This feature is central to the analysis of returns to education, since students who select a particular program may differ systematically in their potential returns from those who select other programs. There is a substantial literature on the empirical content and identification of Roy models, particularly with two sectors; see \cite{HT:2008}, \cite{FT:2011}, and \cite{EHV:2014}. College assignment, however, involves choices among many programs. Only the outcome associated with the attended program is observed, leaving a large number of missing potential outcomes and limiting what can be learned from realized assignments and earnings alone.\footnote{Existing approaches to the identification and estimation of multi-sector selection models impose additional structure or exploit excluded variation; see, for example, \cite{dahl2002}, \cite{heckman2008}, and \cite{heckman2018}.}

Centralized assignment data help address this limitation. Many countries operate such mechanisms, as documented by \cite{kapor2020}. In addition to realized outcomes, these data contain students' submitted rankings and the scores and cutoffs that determine access to programs. Rankings provide information about preferences over alternatives that students do not ultimately attend, while cutoffs distinguish a lack of access from a preference for another program. Under our Roy restriction, these ordinal preferences impose inequalities on potential labor market outcomes. In the baseline specification, students know their potential earnings and rank programs accordingly within a consideration set; nonpecuniary preferences can influence which programs enter that set. For example, ranking one program above another (within a student's consideration set) implies an ordering of potential earnings under the maintained model. An extrapolation that reverses this ordering for students with that ranking would be incompatible with the behavioral restriction, even if it reproduced the evidence at admission cutoffs.

We follow \cite{mourifie2020} and \cite{henry2024} in using partial identification to characterize the information supplied by these restrictions. The assignment mechanism identifies the distribution of realized outcomes conditional on rankings and scores. Continuity identifies additional potential outcome distributions at admission cutoffs. The Roy restriction then translates these identified distributions into bounds on outcomes in other programs. We also consider restrictions on the relationship between admission scores and potential outcomes. In particular, stochastic monotonicity formalizes the hypothesis that higher scores are associated with weakly better potential outcome distributions, conditional on observed characteristics. This restriction connects information across the score distribution and can further tighten the bounds. Related monotonicity restrictions appear in \cite{MP:2009}, \cite{mourifie2020}, and \cite{henry2024}.

Our parameters of interest include average and quantile returns to particular programs, both for the applicant population and for socioeconomic groups. We also study the aggregate and distributional consequences of counterfactual policies, including changes in capacities, admissions priorities, and assignment mechanisms. For a given counterfactual assignment rule, policy effects depend on potential outcomes in the programs to which students would be reassigned. Bounds on potential outcome distributions conditional on rankings and scores therefore translate into bounds on these policy effects. We maintain the invariance of potential outcomes across assignment policies and evaluate counterfactual assignments using the original submitted rankings. The analysis captures changes in the allocation of students across programs while abstracting from peer effects and changes in labor market prices.

We derive closed-form bounds on conditional potential outcome distributions and show how to use them in counterfactual analysis. A numerical illustration considers a reallocation of capacity across programs. The reform changes several admission cutoffs and generates movements between different pairs of programs, making its consequences depend on outcomes beyond a single admission margin. The illustration shows how the Roy restrictions and stochastic monotonicity tighten the bounds relative to those obtained from assignment and continuity alone. It also makes transparent how each source of identifying information contributes to conclusions about the policy.

Our work contributes to the literature on returns to college majors, surveyed by \cite{altonjiarcidiaconomaurel2016}, and connects the literature on local causal effects in education with the structural literature on school choice and assignment mechanisms. The latter includes \cite{agarwal2018}, \cite{fack2019}, \cite{calsamiglia2020}, and \cite{otero2021}, and studies preferences and counterfactual allocations under alternative policies. 

Our framework allows unobserved determinants of preferences and potential outcomes to be dependent on admission scores and, consequently, on students' feasible choice sets. This flexibility matters when scores measure only some of the skills relevant to educational choices and labor market outcomes. Students may possess information about other skills and sort into programs on that basis. Moreover, complementarities between measured and unmeasured skills can generate dependence between scores and unobserved earnings components. These considerations are consistent with the evidence discussed by \cite{borghans2008}, \cite{almlund2011}, and \cite{heckmankautz2012}, and with the skill complementarities studied by \cite{cunha2010}.\footnote{The structural estimates in \cite{arci2005} also point to unobserved heterogeneity affecting admission, financial aid, earnings, and returns to different majors, even conditional on observed test scores.} Allowing such dependence preserves a central feature of selection on comparative advantage.

Finally, the bounds provide a basis for inference using developments in the literature on functional inequalities and stochastic monotonicity, including \cite{Seo:2018}, \cite{LSW:2018}, \cite{HLS:2016}, and \cite{CWK:2021}. We discuss the additional challenges created by conditioning on submitted rankings, including sparse observations for particular lists, and possible approaches to smoothing across rankings.

The remainder of the paper is organized as follows. Section~\ref{sec:method} introduces the assignment framework, parameters of interest, and identifying assumptions. Section~\ref{sec:bounds} derives the bounds for a common scalar admission score and illustrates their implications for counterfactual policy evaluation. Section~\ref{sec:ext} discusses extensions and inference. The appendices collect the proofs and additional results.

\section{General framework and methodology}
\label{sec:method}

\subsection{Model primitives}

We consider a continuum population of high school graduates~$\omega\in\Omega$ assigned to a set of~$J$ university-program pairs (hereafter referred to simply as ``programs''). 

\subsubsection{Student types}

Each high school graduate~$\omega$ is endowed with three objects that constitute their type.
(1) Each student~$\omega$ is endowed with a vector of potential outcomes $(Y_0(\omega),Y_1(\omega),\ldots,Y_J(\omega))$. Each~$Y_j(\omega)$ is the future outcome (for instance, lifetime earnings) of student~$\omega$ if they were randomly assigned to program~$j\in\mJ:=\{1,\ldots,J\}$ or the outside option~$j=0$. The vector of potential outcomes is a primitive of the model: It is unaffected by the assignment of students to programs. Hence, we rule out general equilibrium effects that change the price of talent in a sector of the economy as a result of the assignment of a new cohort of workers. We also rule out peer effects on talents that depend on the assignment of students to programs.
(2) Each student is also endowed with a nonnegative centralized exam grade $S(\omega)$. This may be multidimensional, as with subject tests. The grade~$S(\omega)$ is also a primitive. It rules out choice of effort or test-taking strategy that depends on the assignment mechanism. However, we can evaluate policies that operate through exogenous changes to the grade (``grade subsidies'') for minorities and underrepresented groups. 
(3) Finally, each student is endowed with an information set~$\mathcal I(\omega)$. This is also a primitive. Hence, there is no learning or endogenous information acquisition. However, we can evaluate policies that operate through exogenous changes to the information set such as providing information to applicants. Note that~$\mathcal I(\omega)$ contains~$S(\omega)$. We will make varying assumptions on the information set, including the limiting case of ``perfect foresight,'' where~$\mathcal I(\omega)$ contains the vector of potential outcomes~$(Y_0(\omega),Y_1(\omega),\ldots,Y_J(\omega))$. In what follows, when there is no ambiguity, we shall drop~$\omega$ from the notation.

\subsubsection{Program assignment}

Programs in~$\m{J} := \{1,\ldots, J\}$ have capacities $\{ q_1, \ldots, q_J \}$ defined in terms of shares of the student population. Denote~$\m{J}^0 := \mathcal{J} \cup\{0\}$ to include programs in~$\m{J}$ and the outside option~$0$. We assume there is room for everyone in the outside option, i.e., $q_0=1$. Students are assigned to~$D(\omega)\in\mJ^0$  through a known assignment mechanism that takes as inputs their grade~$S(\omega)$ and a reported rank-order list~$P(\omega)$ of programs in~$\mJ$ that are preferred to the outside option. The list~$P(\omega)$ is submitted by student~$\omega$ on the basis of their information set~$\mathcal I(\omega)$. 
The notation~$jP(\omega)k$ will mean that program~$j$ is listed in $P$ and program~$k$ is either listed in~$P(\omega)$ below $j$ or is the outside option, i.e., $k=0$. 
Throughout, we assume that rank-order lists are sub-orderings of the true preferences of the students reporting them. This rules out a rank-order list~$P(\omega)$ that places program~$k$ ahead of program~$j$ when student~$\omega$ actually prefers~$j$ to~$k$. Student~$\omega$ assigned to program~$D(\omega)$ receives realized outcome~$Y(\omega)=Y_{D(\omega)}(\omega)$.

\subsection{What the analyst observes}

At the student level, we assume that the analyst observes the joint distribution of exam grades~$S$, rank-order lists~$P$, the assigned program $D$, and realized outcomes~$Y_D$ across the student population, in addition to exogenous socioeconomic characteristics we condition the analysis on. 
At the program level, the analyst observes capacities~$(q_1,\ldots,q_J)$ of programs in~$\mJ$ 
and the (nonnegative) deterministic function~$s_j(S)$  of the grade~$S$ used by program~$j$ to evaluate student~$\omega$ based on the specific score $S_j(\omega)=s_j(S(\omega))$.
The observed data allow the analyst to compute the cutoff grade~$c_j$ of program~$j$, defined as the grade of the last admitted student $c_j:=\inf\{S_j: D=j\}$ for all programs~$j\in\mJ$ such that the capacity constraint is binding, and~$0$ otherwise.

\subsection{Parameters of interest}\label{sec:object}

We are interested in causal and counterfactual parameters.

\subsubsection{Causal parameters}

We are interested in the average benefit of program~$j$ relative to program~$k$, i.e.,~$\mathbb E[Y_j-Y_k]$, overall and by socioeconomic group. We are also interested in quantile treatment effects~$Q_{Y_j}(q)-Q_{Y_k}(q)$ for a given quantile~$q$, again overall or by socioeconomic group. In the latter expression, $Q_{Y_j}$ denotes the quantile function of potential outcome~$Y_j$ in program~$j$. Scalar aggregates of quantile treatment effects through rank-dependent social welfare functionals such as~$\int^{1}_{0}\Lambda(q)[Q_{Y_j}(q)-Q_{Y_k}(q)]dq$, for some weight schedule~$\Lambda(q)$, would also be covered in our analysis. Note that we are targeting global parameters, as opposed to treatment effects conditional on grades in the neighborhood of an effective program cutoff, as would be normally obtained with regression discontinuity strategies. 

\subsubsection{Counterfactual parameters}

We consider three types of counterfactual policies and combinations of these: a change in the assignment mechanism;
a change in the grades to equalize the grade distributions of two socioeconomic groups; a change in the program capacities. In all cases, we assume that counterfactual policies are implemented as functions of the original rank-order lists~$P$ submitted by students at the time of the original mechanism. This is a reasonable assumption with strategy-proof mechanisms; see \cite{blm_arxiv} for an identification approach that is robust to strategic reporting. 
We are interested in various functions (primarily the mean and quantiles) of the distribution of counterfactual outcome~$Y^c$ and their differences from the status quo realized outcomes, both in the whole population and within socioeconomic groups. 

Let~$\varphi^c(p,s)$ denote the program to which a student with rank-order list~$P=p$ and grade
~$S=s$ is assigned in the counterfactual assignment mechanism\footnote{The expression is derived for deterministic assignment mechanisms. It extends straightforwardly to cases with extrinsic tie-breaking randomization devices.}. For any function~$g$ of potential outcomes (such as~$g(u):=u$ or~$g(u):=1\{u\leq y\}$), we have
\begin{eqnarray*}
\mathbb E\left[ g(Y^c) \right] & = & \mathbb E\left[ ~ \mathbb E[g(Y^c)\vert P,S] ~ \right] 
\\
& = & \sum_{j\in\mJ^0}\mathbb E\left[~ \mathbb E[\mmi\{\varphi^c(P,S)=j\}g(Y_j)\vert P,S] ~ \right]
\\
& = & \sum_{j\in\mJ^0}\mathbb E\left[~ w^c(j,P,S)\mathbb E[g(Y_j)\vert P,S] ~\right]
\end{eqnarray*}
where the weights~$w^c(j,p,s):=\mmi\{\varphi^c(p,s)=j\}$ are known from the counterfactual assignment mechanism. 
 Under the stability assumption~\ref{ass:stability} of section~\ref{sec:ass}, the status quo assignment is also a function~$p(B(s))$ of~$(p,s)$, where $B(s)$ denotes the set of feasible programs at grade $S=s$ and $p(B(s))$ denotes the most preferred option in that set according to preference $P=p$. 
Thus, the status quo assignment admits the same decomposition as above for the realized outcome~$Y$ with weights~$w(j,p,s):=\mmi\{p(B(s))=j\}$.
Looking at welfare effects of a policy change requires the difference between the counterfactual and the status quo:
\begin{eqnarray}\label{eq:welfare}
\mathbb E\left[g(Y^c)\right] - \mathbb E\left[g(Y)\right] & = & \sum_{j\in\mJ^0}\mathbb E[(w^c(j,P,S)-w(j,P,S))\mathbb E[g(Y_j)\vert P,S]].
\end{eqnarray}
Hence bounds on the conditional mean functions of potential outcomes~$\mathbb E[g(Y_j)\vert P,S]$ allow us to derive bounds on the counterfactual parameters of interest.

\subsection{Identifying assumptions}\label{sec:ass}

We now describe the class of assumptions we make on the endowment distribution, the preferences and the mechanism to derive informative bounds on~$\mathbb E[g(Y_j)\vert P,S]$, hence on the parameters of interest.

\subsubsection{Assumptions on the mechanism}

We assume that the mechanism assigns students to the program in their budget set~$B(S)$ that maximizes their submitted rank-order list~$P$. The budget set~$B(S)$ is defined as the set of programs for which the student's grade~$S$ clears their admission cutoffs, i.e., $B(S):=\{j \in \mJ: s_j(S)\geq c_j\}$. Hence, the realized assignment is~$D=\max P$ in~$B(S)$. 
We will use the notation~$P(B(S))$ for the program that maximizes rank-order list~$P$ in the budget set~$B(S)$.
In case none of the programs in $B(S)$ are listed in $P$, we define $P(B(S))=0$, where $0$ is the outside option. 
Hence, we can write this assumption:
\begin{assumption}[Stability]\label{ass:stability}
$D = P(B(S)).$
\end{assumption}
In the current framework, \cite{azevedo2016} show that stability of the allocation produced by any mechanism is equivalent to $D = P(B(S))$, hence the name of our assumption. 
This allows us to immediately point identify the distribution of~$Y_j$ conditional on~$P=p$ and~$S=s$ for~$j=p(B(s))$. 
Call~$\mathcal D^{stab}$ the set of~$(j,p,s)$ such that the distribution of~$Y_j$ conditional on~$P=p$ and~$S=s$ is point identified in this way. Note that~$(j,p,s) \in \mathcal D^{stab}$ implies~$(l,p,s) \notin \mathcal D^{stab}$ when~$l \neq p(B(s))$. Hence the stability assumption  alone cannot identify causal parameters involving two distinct potential outcomes. Assumption~\ref{ass:stability} can, however, provide restrictions on some counterfactual parameters. More importantly, it helps (partially) identify causal and counterfactual parameters when combined with other assumptions.

\subsubsection{Regularity of preferences and endowments}

\cite{blm_arxiv} derive mild primitive assumptions under which the following continuity holds.
\begin{assumption}[Continuity]\label{ass:continuity} Let $\mathcal{P}(s)$ be the support of the distribution of rank-order lists $P$ conditional on $S=s$.
For any $j \in \mJ^0$, any grade~$s_0$, and $p\in \bigcap_{s \in (s_0-\delta, s_0+\delta)}\mathcal{P}(s)$ for a small $\delta>0$,
the conditional CDF~ 
$
\mmp[ Y_j \leq y \vert P=p, S= s ] 
$
is continuous at $s=s_0$.
\end{assumption}

This assumption allows us to identify the conditional distribution of potential outcome~$Y_k$ 
when $S_j$ is slightly decreased below its cutoff $c_j$ and $k$ becomes the best feasible option.
Formally, define the ``open budget set'' as $B^0(s)=\{j\in \mJ: s_j(s) > c_j \}$.
Whenever $k=p(B^0(s))$ and $k \neq p(B(s))$, continuity implies
\small{\begin{eqnarray*}
\begin{array}{lcl}
\hskip15pt \mmp[ Y_k \leq y \vert P=p, S = s ] 
& = & \lim_{s' \uparrow s} 
\mmp[ Y_k \leq y \vert P=p, S = s' ] 
\\
& = & \lim_{s' \uparrow s} 
\mmp[ Y\leq y \vert P=p, S= s' ].
\end{array}
\end{eqnarray*}
}
Call~$\mathcal D^{cont}$ the set of~$(j,p,s)$ such that the distribution of~$Y_j$ conditional on~$P=p$ and~$S=s$ is point identified in this way.

\subsubsection{Assumptions on preferences}

Only local causal parameters conditional on the grade being equal to or very close to some program cutoffs can be recovered based on Assumptions~\ref{ass:stability} and~\ref{ass:continuity}. Our goal is to recover global causal parameters as well as counterfactual parameters. For this, we strengthen Assumptions~\ref{ass:stability} and~\ref{ass:continuity} with assumptions on preferences and endowments. Our main assumption on preferences is a Roy model assumption adapted to centralized university program assignment environments. We assume that students choose~$P$ in the order of their potential outcomes within a set of programs we call their consideration set~$\mathcal C(P)$. Under Assumptions~\ref{ass:stability} and~\ref{ass:Roy}, they are assigned to an alternative that maximizes their potential earnings among the feasible listed programs and the outside option, i.e., within~$\{0\}\cup(P\cap B(S))$. 

\begin{definition}
    We call~$\mathcal C(\omega)$ the consideration set of student~$\omega$. We assume throughout that two students with the same rank-order list have the same consideration set, and we write~$\mathcal C(\omega)=\mathcal C(P(\omega))$ with slight abuse of notation. We also assume that the programs in the rank-order list itself and the outside option are in the consideration set, i.e., $\{0\} \cup P \subseteq \mathcal C(P) \subseteq \mathcal{J}^0$.
\end{definition}

At one extreme, the consideration set is~$\mathcal J^0$. It may also consist of the outside option together with all the programs in the fields spanned by the rank-order list~$P$, or simply be~$\{0\}\cup P$.

\begin{assumption}[Roy]\label{ass:Roy}
For all~$j\in P$ and $k\in\mathcal C(P)$, ($jPk$ or~$k\notin P)\Rightarrow Y_j\geq Y_k$.
\end{assumption}

This implies that students choose their rank-order list within their consideration set of programs based on labor market future outcomes only, which they have perfect foresight about. Field of study nonpecuniary preferences only affect the choice of consideration set, not the ranking within the latter.
Related work-in-progress by our research team studies the college assignment data matched to future earnings in Norway. 
Figure ~~\ref{fig:monotone} below displays preliminary empirical evidence that supports the Roy assumption in the Norwegian context.

\begin{figure}[htb]
    \begin{subfigure}[b]{0.495\textwidth}  
        \centering
        \includegraphics[width=\textwidth]{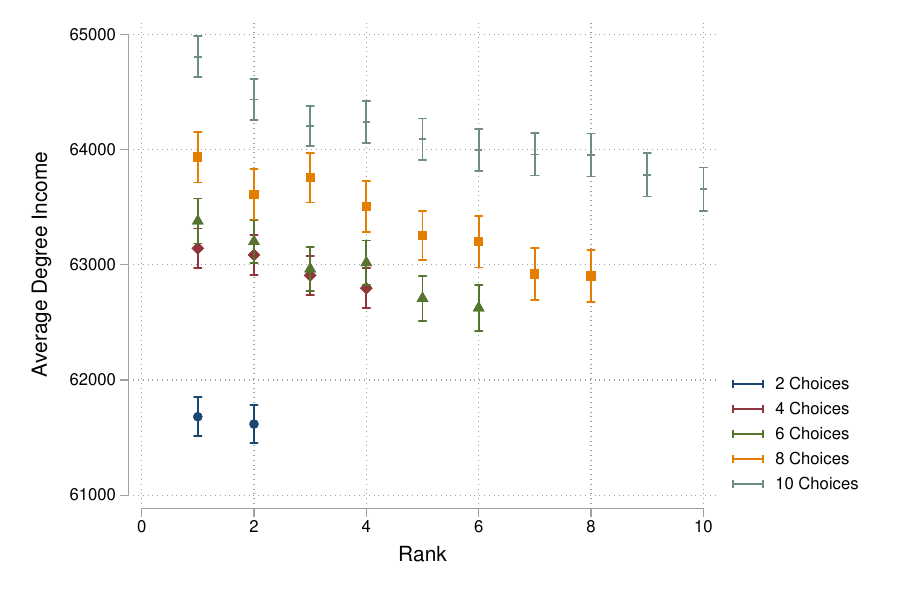}  
        \caption{}
        \label{fig:monotone:female}  
    \end{subfigure}
    \hfill  
    \begin{subfigure}[b]{0.495\textwidth}  
        \centering
        \includegraphics[width=\textwidth]{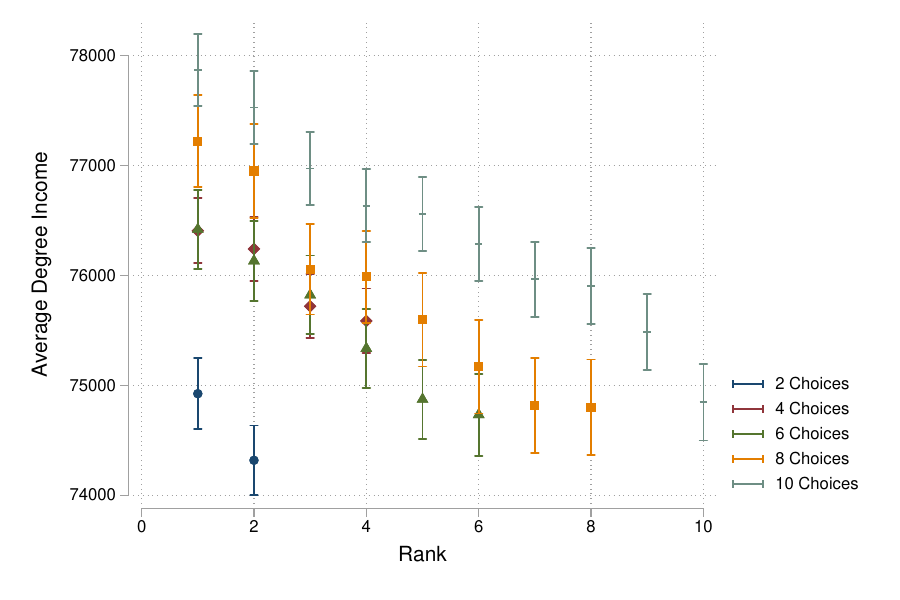}  
        \caption{}
        \label{fig:monotone:male}  
    \end{subfigure}
\vspace{-.4cm}
\caption{\textsc{Preference Rankings and Average Earnings.} \footnotesize{The figure plots average earnings of college--major pairs (henceforth, \textit{degrees}) on the $y$-axis against ranking position on submitted lists on the $x$-axis, using Norwegian centralized college admissions data. The sample of rank-order lists submitted in 2015 is partitioned by gender and list size: panel A shows males and panel B shows females, with colors within each panel denoting list size. Each degree is matched to the average earnings of past graduates of that degree of the same gender as the applicant, computed by averaging winsorized real earnings (1st--99th percentiles, in 2015 USD) over years in which graduates are observed at ages 30--50. For each subsample defined by gender and list size, and for each ranking position, the figure plots the mean and 95\% confidence interval.
}
}
\label{fig:monotone}
\end{figure}

\subsubsection{Assumptions on the type distribution}

The final class of conditions we consider relates to the joint distribution of potential outcomes and grades. Grade~$S_j$ is used by program~$j$ to assess the suitability of a student for a sector. Hence, we shouldn't expect a higher grade~$S_j$ to be systematically associated with worse labor market outcomes. We formalize this with an assumption of stochastic monotonicity of potential outcome~$Y_j$ conditional on~$S_j$.
\begin{assumption}[Monotonicity]
\label{ass:monotonicity}
For all~$j\in\mJ$, all~$s_j\geq s_j'$, and all~$y$,
\begin{eqnarray*}
\mathbb P(Y_j\leq y\vert S_j=s_j) & \leq & \mathbb P(Y_j\leq y\vert S_j=s_j').
\end{eqnarray*}
\end{assumption}
We drop exogenous covariates for ease of notation, but the stochastic monotonicity assumption must be construed as conditional on the latter. Assumption~\ref{ass:monotonicity} is combined with Assumptions~\ref{ass:stability}-\ref{ass:Roy} to derive bounds on the causal and counterfactual parameters of interest in the following way. From assumption~\ref{ass:monotonicity}, we obtain for all~$j\in\mJ$ and all~$y$,
\begin{eqnarray}
\label{eq:mon}
\begin{array}{lclcl}
\sup_{s_j'\geq s_j}\mathbb P(Y_j\leq y\vert S_j=s_j') & \leq & \mathbb P(Y_j\leq y\vert S_j=s_j) & \leq & \inf_{s_j'\leq s_j}\mathbb P(Y_j\leq y\vert S_j=s_j'). 
\end{array}
\end{eqnarray}
Recall that~$\mathcal P(s)$ is the support of rank-order lists~$P$ conditional on~$S=s$. For instance, rank-order lists submitted by students with high grades may exclude programs that these students rank below the outside option. We have
\begin{eqnarray}
\label{eq:split}
\mathbb P(Y_j\leq y\vert S_j=t)
&=&\int \sum_{p\in\mathcal P(s)}\mathbb P(P=p\vert S=s)
\;\mathbb P(Y_j\leq y\vert P=p,S=s)\,dF_{S\mid S_j=t}(s).
\end{eqnarray}
Here~$F_{S\mid S_j=t}$ is the conditional distribution of the full grade vector~$S$ given~$S_j=t$; it is supported on~$\{s:s_j(s)=t\}$. This expression accommodates both discrete and continuous grades. When~$S_j=S$ is a common scalar grade, it reduces to the sum over rank-order lists at~$S=t$.

Combining~(\ref{eq:mon}) and~(\ref{eq:split}) with the bounds~(\ref{eq:roy-bounds1})-(\ref{eq:roy-bounds2}) on~$\mathbb P(Y_j\leq y\vert P=p,S=s)$ derived from Assumptions~\ref{ass:stability}-\ref{ass:Roy} in section~\ref{sec:bounds} yields additional constraints on~$\mathbb P(Y_j\leq y\vert P=p,S=s)$. This suggests an iterative procedure; Proposition~\ref{prop:iteration} in the appendix shows that a single iteration exhausts the information combined in this way.

\subsection{Bringing it all together}

We obtain informative bounds on the causal and counterfactual parameters described in section~\ref{sec:object} by optimization under constraints on~$\mathbb P(Y_j\leq y\vert P=p,S=s)$ obtained under Assumptions~\ref{ass:stability}-\ref{ass:monotonicity}. It is important to bound the CDFs conditional on the rank-order list~$P$ for two reasons. First, it allows us to consider counterfactual policies that are functions of the rank-order list~$P$ in section~\ref{sec:object}. Second, it allows us to fully exploit Assumptions~\ref{ass:stability}-\ref{ass:monotonicity}. 
Section~\ref{sec:bounds} derives closed-form bounds for~$\mathbb P(Y_j\leq y\vert P=p,S=s)$ when all programs use a unique scalar grade.

\section{Bounds with common scalar grade}
\label{sec:bounds}

This section is devoted to a complete derivation of the bounds on~$\mathbb P(Y_j\leq y\vert P=p,S=s)$ in closed form and a simulation experiment in case the students take a unique exam, which is used by all programs. Throughout the section, the grade~$S$ is therefore scalar and~$S_j=S$ for all~$j\in\mJ$, so that~$B(s)=\{j\in\mJ: s\geq c_j\}$ and~$B^0(s)=\{j\in\mJ: s>c_j\}$. Since assumption~\ref{ass:Roy} restricts realized potential outcomes rather than their conditional expectations, students are implicitly endowed with perfect foresight: their information set contains the potential labor market outcomes.

We maintain the following conventions. (i) $\{0\}\cup P\subseteq\mathcal C(P)\subseteq\mJ^0$: students consider the programs they list and the outside option, as in the examples of consideration sets given in section~\ref{sec:method}. (ii) The outside option is treated as a considered, unlisted alternative: assumption~\ref{ass:Roy} with~$k=0$ gives~$Y_j\geq Y_0$ for all~$j\in P$, which formalizes the maintained hypothesis of section~\ref{sec:method} that listed programs are preferred to the outside option; assumption~\ref{ass:Roy} does not otherwise restrict~$Y_0$. (iii) $p(B(s))=0$ when no program listed in~$p$ belongs to~$B(s)$, and similarly for~$p(B^0(s))$; by assumption~\ref{ass:stability}, students with~$P=p$ and~$S=s$ such that $p(B(s))=0$ are then assigned to the outside option. Finally, for~$k\in\mJ^0$, we write
\[
F_k(y\vert p,s):=\mathbb P[Y_k\leq y\vert P=p,S=s]
\]
for the conditional CDF of potential outcome~$Y_k$, which is the object we bound.

\subsection{Bounds on the potential outcome distributions}

\subsubsection{Point identification}

Assumptions~\ref{ass:stability} and~\ref{ass:continuity} point identify~$F_k(y\vert p,s)$ for up to two alternatives at each~$(p,s)$: the alternative~$p(B(s))$ students are assigned to, and the alternative~$p(B^0(s))$ they would be assigned to if their grade were marginally lower.
\begin{eqnarray}\label{eq:pointid}
\begin{array}{lcll}
F_{p(B(s))}(y\vert p,s) & = & \mmp[ Y \leq y \vert P=p, S= s ], & \\
F_{p(B^0(s))}(y\vert p,s) & = & \lim_{s' \uparrow s}\mmp[ Y\leq y \vert P=p, S= s' ] & \mbox{ when } p(B^0(s))\ne p(B(s)).
\end{array}
\end{eqnarray}
The second case arises when~$s$ is the cutoff of the assigned program, i.e.,~$s=c_{p(B(s))}$. It includes~$p(B^0(s))=0$, the case where~$s$ is the cutoff of the last listed program in the budget set, in which assumption~\ref{ass:continuity} identifies the distribution of the outside option outcome~$Y_0$. When~$p(B^0(s))=p(B(s))$, only one potential outcome distribution is identified at~$(p,s)$. The sets~$\mathcal D^{stab}$ and~$\mathcal D^{cont}$ of section~\ref{sec:method} collect the corresponding triplets~$(j,p,s)$.

\subsubsection{Bounds from the Roy assumption}

Extend the ordering of the list~$p$ to~$\mJ^0$: for~$j,k\in\mJ^0$, write~$j\succ_p k$ if~$j\in p$, $k\in\mathcal C(p)\cup\{0\}$, and either~$j\,p\,k$ or~$k\notin p$; write~$j\succeq_p k$ if~$j\succ_p k$ or~$j=k$. In this notation, assumption~\ref{ass:Roy} states that for students with~$P=p$, $j\succ_p k\Rightarrow Y_j\geq Y_k.$ Fix~$(p,s)$ and~$j\in\mJ^0$. Assumption~\ref{ass:Roy} thus bounds~$F_j(y\vert p,s)$ above by the identified CDF of any alternative ranked weakly below~$j$, and below by the identified CDF of any alternative ranked weakly above~$j$. Define
\begin{eqnarray}\label{eq:roy-bounds1}
U_j(y\vert p,s) & := & \min\left\{F_k(y\vert p,s):\; k\in\{p(B(s)),p(B^0(s))\},\; j\succeq_p k\right\},
\end{eqnarray}
and
\begin{eqnarray}\label{eq:roy-bounds2}
L_j(y\vert p,s) & := & \max\left\{F_k(y\vert p,s):\; k\in\{p(B(s)),p(B^0(s))\},\; k\succeq_p j\right\},
\end{eqnarray}
with the conventions~$\min\emptyset:=1$ and~$\max\emptyset:=0$. Both bounds are identified under Assumptions~\ref{ass:stability} and~\ref{ass:continuity} by~(\ref{eq:pointid}), and Theorem~\ref{thm:order} below gives~(\ref{eq:roy-bounds1}) and~(\ref{eq:roy-bounds2}) in closed form and establishes their validity under Assumptions~\ref{ass:stability}-\ref{ass:Roy}. 
\begin{theorem}\label{thm:order}
Under Assumptions~\ref{ass:stability}-\ref{ass:Roy}, for all~$j\in\mJ^0$, $y$, $s$ and~$p\in\m{P}(s)$, we have the bounds $L_j(y\vert p,s)\leq F_j(y\vert p,s)\leq U_j(y\vert p,s)$, with:
\begin{eqnarray*}
\begin{array}{lllll}
U_j(y\vert p,s) 
& = &
F_{p(B(s))}(y\vert p,s) &\textup{ if }& j \succeq_p p(B(s)),\\
& = &
F_{p(B^0(s))}(y\vert p,s) &\textup{ if }& p(B(s)) \succ_p j \succeq_p p(B^0(s)),\\
& = &
1 &\textup{ otherwise},&
\end{array}
\end{eqnarray*}
and
\begin{eqnarray*}
\begin{array}{lllll}
L_j(y\vert p,s) 
& = &
F_{p(B^0(s))}(y\vert p,s) &\textup{ if }& p(B^0(s)) \succeq_p j,\\
& = &
F_{p(B(s))}(y\vert p,s) &\textup{ if }& p(B(s)) \succeq_p j \textup{ and } p(B^0(s))\not\succeq_p j,\\
& = &
0 &\textup{ otherwise}.&
\end{array}
\end{eqnarray*}
\end{theorem}
When~$p(B^0(s))=p(B(s))$, the second case of each display in Theorem~\ref{thm:order} is empty, and the bounds only involve the CDF of the realized outcome. When~$p(B^0(s))\ne p(B(s))$, the three cases for~$U_j$ correspond to~$j$ ranked weakly above the assigned program~$p(B(s))$, to~$j$ ranked strictly below~$p(B(s))$ and weakly above the program~$p(B^0(s))$ identified by continuity, and to~$j$ not ranked weakly above~$p(B^0(s))$, which includes every unlisted program when~$p(B^0(s))=0$. The cases for~$L_j$ run in the opposite direction: $j$ ranked weakly below~$p(B^0(s))$; $j$ ranked weakly below~$p(B(s))$ but not weakly below~$p(B^0(s))$; and~$j$ not ranked weakly below~$p(B(s))$, which includes the programs listed above~$p(B(s))$ and the programs outside~$\mathcal C(p)$.

\subsubsection{Bounds from monotonicity}

From assumption~\ref{ass:monotonicity}, we have the constraints~(\ref{eq:mon}), with~$S_j=S$, for any~$j\in\mJ$, $s$ and~$y$.
These constraints apply to the CDFs conditional only on $S$, instead of both $S$ and $P$ as required.
We convert them into bounds on~$F_j(y\vert p,s)$ in the following way.
Fix $s$ and $p \in \m{P}(s)$.
Define 
\[
w^{p'\vert s'}_{p\vert s}:= \frac{\mmp[P=p' \vert S=s' ]}{\mmp[P=p \vert S=s ]}.
\]
Writing~$\mmp[Y_j\leq y\vert S=s]=\sum_{p' \in \m{P}(s)} 
\mmp[P=p' \vert S=s ] F_j(y\vert p',s)$ as the sum of~$\mmp[P=p \vert S=s ] F_j(y\vert p,s)$ and the residual sum over~$p'\ne p$ immediately yields
\begin{eqnarray*}
F_j(y\vert p,s) & = &
\frac{\mmp[ Y_j \leq y \vert S=s ]}{\mmp[P=p \vert S=s ]}
-
\sum_{\stackrel{p' \in \m{P}(s)}{p' \neq p}}
w^{p'\vert s}_{p\vert s} \;
F_j(y\vert p',s)  .
\end{eqnarray*}
The first term is bounded by~(\ref{eq:mon}), where, for each~$s'$, $\mmp[Y_j\leq y\vert S=s']$ is in turn bounded by the average over~$p'\in\m{P}(s')$ of the Roy bounds~(\ref{eq:roy-bounds1}) and~(\ref{eq:roy-bounds2}) at~$(p',s')$. Each term of the residual sum is bounded by the Roy bounds at~$(p',s)$. This motivates the following definitions:
\begin{equation}\label{eq:mon-bounds}
\begin{aligned}
\tilde U_j(y\vert p,s) & :=
\inf_{s'\leq s } \left( \sum_{p' \in \m{P}(s')} 
w^{p'\vert s'}_{p\vert s} \;
U_j(y\vert p',s') \right)
-
\sum_{\stackrel{p' \in \m{P}(s)}{p' \neq p}}
w^{p'\vert s}_{p\vert s} \;
L_j(y\vert p',s),
\\
\tilde L_j(y\vert p,s) & :=
\sup_{s'\geq s } \left( \sum_{p' \in \m{P}(s')} 
w^{p'\vert s'}_{p\vert s} \;
L_j(y\vert p',s') \right)
-
\sum_{\stackrel{p' \in \m{P}(s)}{p' \neq p}}
w^{p'\vert s}_{p\vert s} \;
U_j(y\vert p',s).
\end{aligned}
\end{equation}
The following theorem, proved in Appendix~\ref{app:proofs}, collects the bounds on~$F_j(y\vert p,s)$ obtained from Assumptions~\ref{ass:stability}-\ref{ass:monotonicity}.
\begin{theorem}\label{thm:bounds}
Under Assumptions~\ref{ass:stability}-\ref{ass:monotonicity}, for all~$j\in\mJ$, $y$, $s$ and~$p\in\m{P}(s)$, with~$\tilde L_j$ and~$\tilde U_j$ defined in~(\ref{eq:mon-bounds}), we have the bounds
\begin{eqnarray}\label{eq:final}
\begin{array}{lclcl}
\max\{L_j(y\vert p,s),\tilde L_j(y\vert p,s)\} & \leq & F_j(y\vert p,s) & \leq & \min\{U_j(y\vert p,s),\tilde U_j(y\vert p,s)\}.
\end{array}
\end{eqnarray}

\end{theorem}
Substituting the bounds in~(\ref{eq:final}) for~$L_j$ and~$U_j$ in the definitions of~$\tilde L_j$ and~$\tilde U_j$ suggests an iterative tightening. Proposition~\ref{prop:iteration} in the appendix shows that a single iteration exhausts the information in~(\ref{eq:mon}) and~(\ref{eq:roy-bounds1})-(\ref{eq:roy-bounds2}) combined in this way. Thus, further iterations will not refine \eqref{eq:final}.

\subsection{Sharp bounds on counterfactual parameters}
\label{sec:cf}

We now turn to the welfare parameter~(\ref{eq:welfare}) when the counterfactual assignment~$\varphi^c$ is a known function of~$(P,S)$, so that the weights~$w(j,p,s)=\mmi\{p(B(s))=j\}$ and~$w^c(j,p,s)=\mmi\{\varphi^c(p,s)=j\}$ in~(\ref{eq:welfare}) are fixed. We take~$g(u):=\mmi\{u\leq y\}$ for a fixed~$y$, so that~(\ref{eq:welfare}) is the distributional effect
\begin{eqnarray}\label{eq:theta}
\theta(y) & := & \mmp[Y^c\leq y]-\mmp[Y\leq y] \;=\; \mme\left[F_{\varphi^c(P,S)}(y\vert P,S)-F_{P(B(S))}(y\vert P,S)\right],
\end{eqnarray}
where the second equality uses~$Y^c=Y_{\varphi^c(P,S)}$ and assumption~\ref{ass:stability}. Write~$\pi_s(p):=\mmp[P=p\vert S=s]$, so that~$w^{p'\vert s'}_{p\vert s}=\pi_{s'}(p')/\pi_s(p)$. For~$j\in\mJ^0$, call
\[
\mathcal M_j(s):=\{p\in\m{P}(s):\;\varphi^c(p,s)=j\neq p(B(s))\}
\]
the set of lists moved into~$j$ at grade~$s$ by the policy. Lists that are not moved contribute nothing to~(\ref{eq:theta}), so that
\begin{eqnarray}\label{eq:theta-movers}
\theta(y) & = & \mme\Big[\sum_{j\in\mJ^0}\;\sum_{p\in\mathcal M_j(S)}\pi_S(p)\left(F_j(y\vert p,S)-F_{p(B(S))}(y\vert p,S)\right)\Big],
\end{eqnarray}
where the expectation is with respect to~$S$. In~(\ref{eq:theta-movers}), $F_{p(B(s))}(y\vert p,s)$ is identified by~(\ref{eq:pointid}) and~$F_j(y\vert p,s)$ is the unknown.

\begin{theorem}\label{thm:cf}
Under Assumptions~\ref{ass:stability} and~\ref{ass:Roy}, and the implications~(\ref{eq:pointid}) of assumption~\ref{ass:continuity}, the sharp\footnote{The bounds in Theorem~\ref{thm:cf} are pointwise sharp, i.e., sharp for each value of the outcome~$y$. They are not functionally sharp bounds.} identified set for~$\theta(y)$ is~$[\underline\theta^R(y),\overline\theta^R(y)]$, with
\begin{eqnarray*}
\underline\theta^R(y) & := & \mme\Big[\sum_{j\in\mJ^0}\;\sum_{p\in\mathcal M_j(S)}\pi_S(p)\left(L_j(y\vert p,S)-F_{p(B(S))}(y\vert p,S)\right)\Big],\\
\overline\theta^R(y) & := & \mme\Big[\sum_{j\in\mJ^0}\;\sum_{p\in\mathcal M_j(S)}\pi_S(p)\left(U_j(y\vert p,S)-F_{p(B(S))}(y\vert p,S)\right)\Big].
\end{eqnarray*}
\end{theorem}
The identified set in Theorem~\ref{thm:cf} is the set of values of~$\theta(y)$ generated by models that satisfy Assumptions~\ref{ass:stability}-\ref{ass:Roy} and reproduce the distribution of~$(P,S,Y)$. The proof in Appendix~\ref{app:proofs} shows that it is the image of the convex set of families of conditional CDFs that satisfy~(\ref{eq:pointid}) and~$j\succeq_p k\Rightarrow F_j(\cdot\vert p,s)\leq F_k(\cdot\vert p,s)$, hence an interval, and that the bounds of Theorem~\ref{thm:order} are attained by such families simultaneously in all cells~$(p,s)$, since Assumptions~\ref{ass:stability}-\ref{ass:Roy} do not link the cells. Bounds on quantiles of~$Y^c$ follow by inversion and, for continuous nondecreasing~$g$ and bounded outcomes, (\ref{eq:welfare}) equals~$-\int\theta(y)\,dg(y)$, so that Theorem~\ref{thm:cf} also delivers bounds on~(\ref{eq:welfare}). When assumption~\ref{ass:monotonicity} is added, substituting~(\ref{eq:final}) for~$L_j$ and~$U_j$ in Theorem~\ref{thm:cf} yields valid bounds on~$\theta(y)$, which are not sharp in general, since assumption~\ref{ass:monotonicity} links the cells of each program across lists, grades and values of~$y$; the sharp bounds are then the values of a linear program over families of conditional CDFs that satisfy~(\ref{eq:pointid}), the Roy ordering and the monotonicity of~$s\mapsto\mmp[Y_j\leq u\vert S=s]$ for all~$u$.

\subsection{Numerical illustration}

This section illustrates, in a stylized case with three programs, how the bounds on~$\mathbb P(Y_j\leq y\vert P=p,S=s)$ are derived and how informative the resulting bounds on counterfactual parameters are.


\subsubsection{Description of the data generating process}

There are~$J$ programs, each with capacity~$q_j=1/J$, $j=1,\ldots,J$. The distribution of student grades~$S$ is uniform on~$[0,1]$. The distribution of the vector of potential outcomes~$(Y_1,\ldots,Y_J)$ conditional on~$S=s$ is multivariate lognormal. The underlying multivariate normal distribution has mean~$(\alpha_1s,\ldots,\alpha_Js)^\top$,
with positive weights~$\alpha_j$ such that~$\alpha_1>\ldots>\alpha_J$, and identity variance-covariance matrix. The outside option outcome is~$Y_0=0$, so that every program is preferred to the outside option.


\subsubsection{Allocation of students to programs}

There is a continuum of students~$\omega\in\Omega$. Each student~$\omega$ is endowed with a grade~$S(\omega)$ drawn independently from~$U[0,1]$, and a vector of potential outcomes~$(Y_{1}(\omega),\ldots,Y_{J}(\omega))$ drawn from the lognormal distribution described above. Each student~$\omega$ submits a rank-order list~$P(\omega)$ of programs~$j\in\mJ$ in the order of potential outcomes~$Y_{j}(\omega)$. Lists are complete, so that~$P=\mJ$ and~$\mathcal C(P)=\mJ^0$. Since~$Y_0=0$ and all program outcomes are strictly positive, assumption~\ref{ass:Roy} holds. The extended order~$\succ_P$ coincides with the order of the list on~$\mJ$ and ranks every listed program above the outside option. The bounds of section~\ref{sec:bounds} therefore do not depend on the consideration set in this illustration; truncated lists or smaller consideration sets would widen them through the trivial Roy bounds on unconsidered programs. Allocation of students to programs follows serial dictatorship based on the grade~$S$: Rank students in decreasing order of grade~$S(\omega)$. Each student~$\omega$ is assigned to the program with the highest~$Y_{j}$ among programs that still have open seats. Call~$c_j$ the cutoff grade for program~$j$, i.e., the lowest grade among all students allocated to program~$j$. Since capacities add up to one, every student is assigned to a program and the outside option plays no role.

\subsubsection{Bounds on potential outcome distributions in the case with~$3$ programs.}

We consider first a case with 3 programs: $J=3$, $q_1=q_2=q_3=1/3$, $\alpha_1=12,$ $\alpha_2=11$ and~$\alpha_3=10$. We derive cutoffs~$c_1=0.4788,c_2=0.1474,c_3=0$, compile the list of triplets~$(j,p,s)\in\mathcal D^{stab}\cup\mathcal D^{cont}$ (i.e., triplets~$(j,p,s)$ such that the CDF of potential outcome~$Y_j$ conditional on~$P=p,S=s$ is point identified), and finally compute the bounds on the CDF of~$Y_j\vert P=p,S=s$ for the remaining~$(j,p,s)$. For the latter, we compare bounds under Assumptions~\ref{ass:stability}-\ref{ass:Roy} in~(\ref{eq:roy-bounds1}) and~(\ref{eq:roy-bounds2}) and the tighter bounds in~(\ref{eq:final}) obtained when assumption~\ref{ass:monotonicity} is added.

Table~\ref{table:D} lists all the triplets~$(j,p,s)\in\mathcal D^{stab}\cup\mathcal D^{cont}$, i.e., such that~$y\mapsto \mathbb P(Y_j\leq y\vert P=p,S=s)$ 
is point identified from the stability assumption~\ref{ass:stability} or the continuity assumption~\ref{ass:continuity}. For instance, if the rank-order list is~$p=(1,2,3)$ and the grade is~$s\in[c_2,c_1)$, then~$Y_D=Y_2$, so that~$y\mapsto \mathbb P(Y_2\leq y\vert P=(1,2,3),S=s)$ is point identified and~$(2,p,s)\in\mathcal D^{stab}$. If the rank-order list is~$p=(1,2,3)$ and the grade is~$s=c_1$, then~$y\mapsto \mathbb P(Y_2\leq y\vert P=(1,2,3),S=c_1)$ is point identified by continuity as~$s'\uparrow s=c_1$ and~$(2,p,s)\in\mathcal D^{cont}$.\footnote{Table~\ref{table:D} does not explicitly consider 
randomly generated priority scores (e.g., lottery tie breakers), which identifies more triplets $(j,p,s)$ than currently listed in Table~\ref{table:D}. See, for example, Section 3.1 of \cite{chen2023nonparametric}.
}

\begin{table}
\begin{center}
\begin{tabular}{c|clclc}
$P$ & $S\in[c_3,c_2)$ & $S=c_2$ & $S\in(c_2,c_1)$ & $S=c_1$ & $S\in(c_1,1]$ \\ \hline
$1\;2\;3$ & $Y_3$ & $Y_2\;(Y_3)$ & $Y_2$ & $Y_1\;(Y_2)$ & $Y_1$ \\
$1\;3\;2$ & $Y_3$ & $Y_3$ & $Y_3$ & $Y_1\;(Y_3)$ & $Y_1$ \\
$2\;1\;3$ & $Y_3$ & $Y_2\;(Y_3)$ & $Y_2$ & $Y_2$ & $Y_2$ \\
$2\;3\;1$ & $Y_3$ & $Y_2\;(Y_3)$ & $Y_2$ & $Y_2$ & $Y_2$ \\
$3\;1\;2$ & $Y_3$ & $Y_3$ & $Y_3$ & $Y_3$ & $Y_3$ \\
$3\;2\;1$ & $Y_3$ & $Y_3$ & $Y_3$ & $Y_3$ & $Y_3$ 
\end{tabular}
\end{center}
\caption{Table of point-identified CDFs. If potential outcome~$Y_j$ appears in cell~$P\times S$, it means~$Y_j\vert P,S$ is identified by stability. If it appears in parentheses, it is identified by continuity.}
\label{table:D}
\end{table}

\begin{figure}
\centering
	\includegraphics[width=0.45\linewidth]{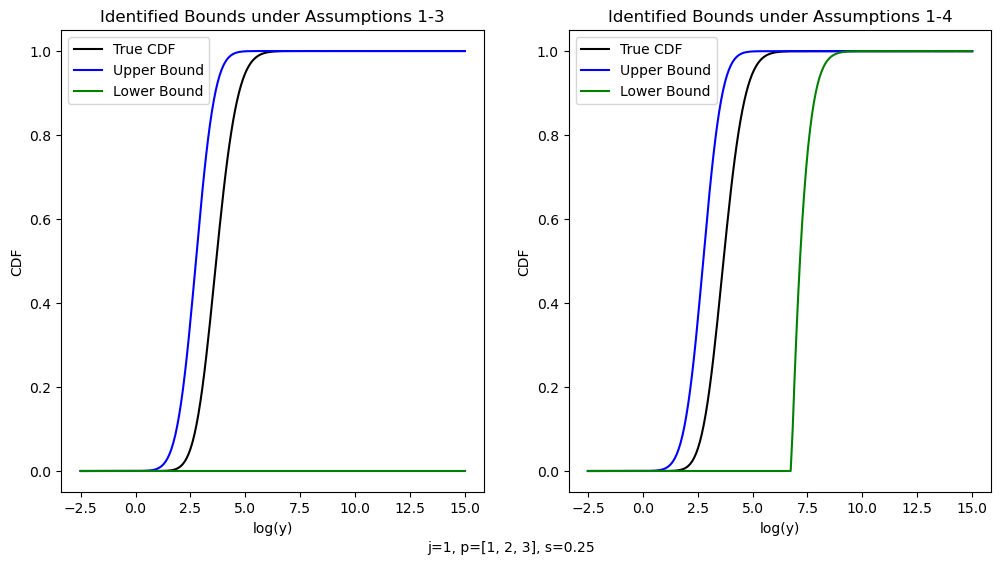}
	\includegraphics[width=0.45\linewidth]{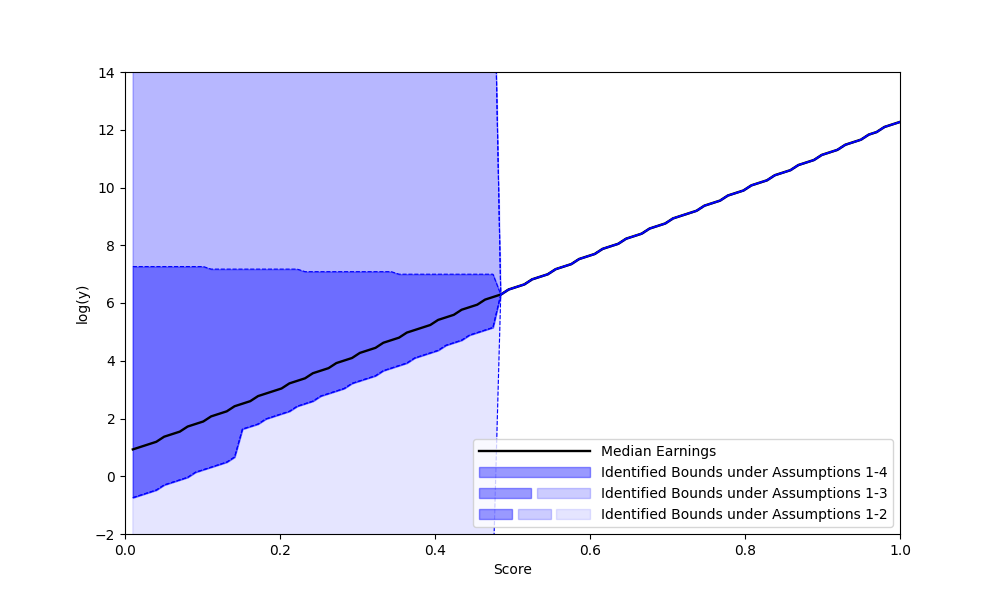}
\caption{Left panel (first and second plots): true CDF of~$Y_1$ conditional on~$P=(1,2,3)$ and~$S=0.25$, bounds under Assumptions~\ref{ass:stability}-\ref{ass:Roy} from~(\ref{eq:roy-bounds1}) and~(\ref{eq:roy-bounds2}), and bounds under Assumptions~\ref{ass:stability}-\ref{ass:monotonicity} from~(\ref{eq:final}). Right panel (third plot): bounds on the conditional median of~$Y_1$ given~$P=(1,2,3)$ and~$S=s$, obtained by inverting the bounds on the conditional CDF, as functions of~$s$.}
\label{fig:CDF}
\end{figure}

The left panel of figure~\ref{fig:CDF} shows the true CDF of potential outcome~$Y_1$ conditional on~$P=(1,2,3)$ and~$S=0.25$, together with the lower and upper bounds under Assumptions~\ref{ass:stability}-\ref{ass:Roy} from~(\ref{eq:roy-bounds1}) and~(\ref{eq:roy-bounds2}), and the bounds under Assumptions~\ref{ass:stability}-\ref{ass:monotonicity} from~(\ref{eq:final}), of which the lower bound is tighter. The right panel of figure~\ref{fig:CDF} shows the bounds on the conditional median of~$Y_1$ given~$P=(1,2,3)$ and~$S=s$, obtained by inverting the bounds on the conditional CDF, as functions of the grade~$s$ ranging from~$0$ to~$1$.

\subsubsection{Bounds on counterfactual parameters}

\begin{table}
\begin{center}
\begin{tabular}{c|cccc|cccc}
 & \multicolumn{4}{c}{$s\in[c_2,c_2^c)$} & \multicolumn{4}{c}{$s\in[c_1^c,c_1)$} \\ 
$p$ & $j=$ & 3 & 2 & 1 & $j=$ & 3 & 2 & 1 \\ \hline
$1\;2\;3$ & & 1 & -1 & 0 & & 0 & -1 & 1 \\
$1\;3\;2$ &  & 0 & 0 & 0 & & -1 & 0 & 1 \\
$2\;1\;3$ &  & 1 & -1 & 0 & & 0 & 0 & 0 \\
$2\;3\;1$ &  & 1 & -1 & 0 & & 0 & 0 & 0
\end{tabular}
\end{center}
\caption{Table of weights~$w^c(j,p,s)-w(j,p,s)$, which are~$1$ for moves into~$j$, $-1$ for moves out of~$j$ and~$0$ otherwise. The weights are all~$0$ for values of~$(j,p,s)$ not listed in the table.}
\label{table:W}
\end{table}

\begin{figure}
\centering
\includegraphics[width=0.47\linewidth]{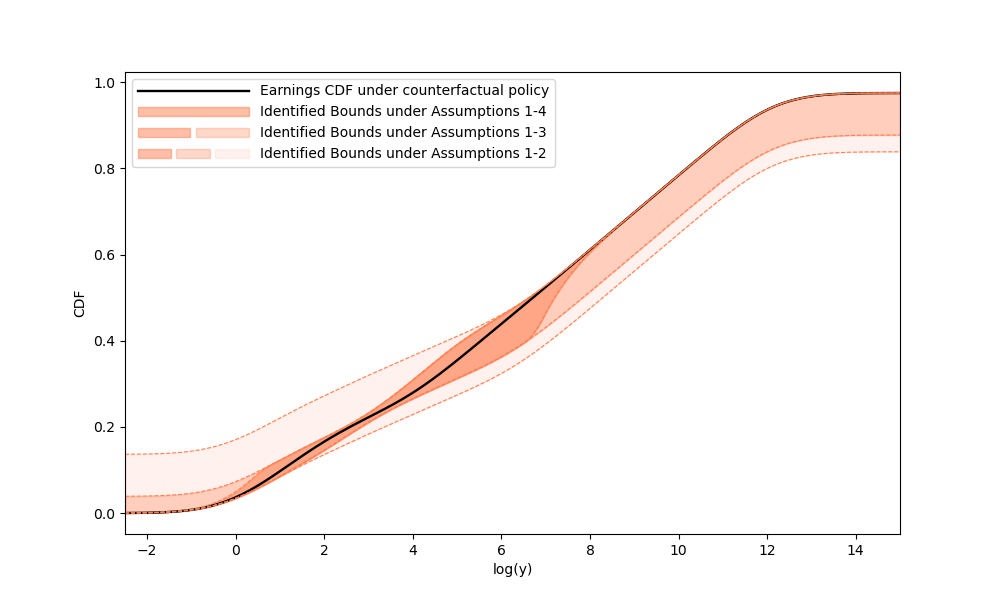}
\includegraphics[width=0.47\linewidth]{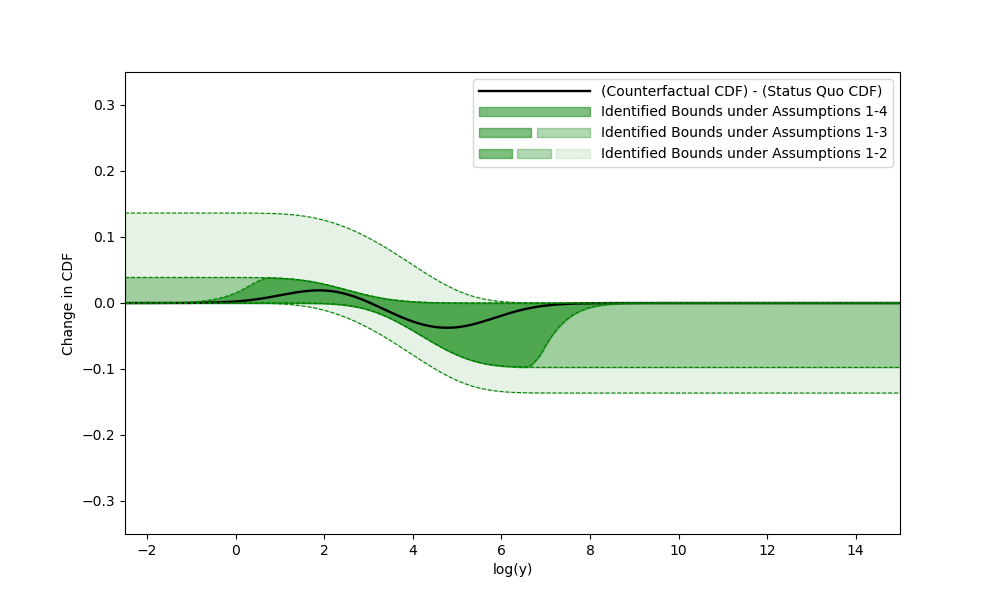}
\caption{Capacity change~$(1/3+\delta,1/3-\delta,1/3)$ with~$\delta=0.1$. Left panel: true CDF of~$Y^c$ and bounds under Assumptions~\ref{ass:stability}-\ref{ass:continuity} only, under Assumptions~\ref{ass:stability}-\ref{ass:Roy} from~(\ref{eq:roy-bounds1}) and~(\ref{eq:roy-bounds2}), and under Assumptions~\ref{ass:stability}-\ref{ass:monotonicity} from~(\ref{eq:final}). Right panel: true~$\theta(y)=\mathbb P(Y^c\leq y)-\mathbb P(Y\leq y)$ and the corresponding bounds.}
\label{fig:counter}
\end{figure}

Figure~\ref{fig:counter} shows the bounds on the counterfactual aggregate CDF of realized outcomes~$Y_D$ after a change in program capacities. The counterfactual policy entertained is the following. Program capacities are changed to~$(q_1=1/3+\delta,q_2=1/3-\delta,q_3=1/3)$ for
~$\delta=0.1$. Students report the same rank-order lists and are allocated according to the same procedure. The cutoff of program~$1$ goes down from~$c_1=0.4788$ to~$c_1^c=0.2777$, while the cutoff for program~$2$ goes up from~$c_2=0.1474$ to~$c^c_2=0.2224$. As a result, students with grade~$s\in[c_1^c,c_1)$ who list program~$1$ first move to program~$1$ (from program~$2$ if~$2p3$, from program~$3$ otherwise), and students with grade~$s\in[c_2,c_2^c)$ and~$2p3$ move from program~$2$ to program~$3$; see Table~\ref{table:W}. Let~$Y^c$ be the outcome under the counterfactual policy. For illustrative purposes, suppose the welfare measure we consider is a quantile of the outcome distribution. The effect of the policy on the distribution of outcomes is then~$\theta(y)=\mathbb P(Y^c\leq y)-\mathbb P(Y\leq y)=\sum_{j\in\{1,2,3\}}\mathbb E[(w^c(j,P,S)-w(j,P,S))F_j(y\vert P,S)]$, which is linear in the unknown conditional CDFs. Bounds on~$\theta(y)$ are obtained by substituting cell by cell bounds on~$F_j(y\vert p,s)$ into this expression: the bounds~$0\leq F_j(y\vert p,s)\leq1$ on the CDFs that are not point identified under Assumptions~\ref{ass:stability} and~\ref{ass:continuity} only; the Roy bounds~(\ref{eq:roy-bounds1}) and~(\ref{eq:roy-bounds2}) under Assumptions~\ref{ass:stability}-\ref{ass:Roy}, which yield the sharp bounds of Theorem~\ref{thm:cf}; and the bounds~(\ref{eq:final}) under Assumptions~\ref{ass:stability}-\ref{ass:monotonicity}, which are valid but not sharp in general, as noted after Theorem~\ref{thm:cf}. The left panel of figure~\ref{fig:counter} displays the true CDF of~$Y^c$ together with the three sets of bounds, and the right panel displays the true~$\theta(y)$ together with the corresponding bounds.

\section{Discussion and extensions}
\label{sec:ext}

We propose several extensions of the methodology in ongoing work.

\subsection{Multiple potential outcomes}

Researchers may be interested in the impacts of matching assignment on various outcomes at the same time, e.g., labor market outcomes, nonpecuniary benefits, 
and measures of satisfaction. This is desirable for two principal reasons. First, data on multiple outcomes are often available. Existing approaches treat each potential outcome independently. This disregards dependence between potential outcomes, which can tighten bounds on counterfactual parameters, even when it is not of interest in its own right. Second, when the potential outcomes are labor market earnings or similar pecuniary measures, assumption~\ref{ass:Roy} implies that students make decisions based on pecuniary incentives only. Considering multiple outcomes, including labor market nonpecuniary benefits or measures of satisfaction, relaxes the strong and unrealistic assumption of purely pecuniary incentives, while preserving the empirical content of assumption~\ref{ass:Roy}. The latter would be replaced by an assumption that requires program~$j$ to be listed above program~$k$ if both are listed and~$Y_j\geq Y_k$ for some ordering~$\geq$ on~$\mathbb R^d$. The minimal ordering is the componentwise order on~$\mathbb R^d$. Stronger partial (or complete) orderings can be chosen, depending on the context.

The identifying assumption that needs to be adjusted to accommodate multiple potential outcomes is assumption~\ref{ass:monotonicity}. When potential outcome~$Y_j$ is a vector, assumption~\ref{ass:monotonicity} can be extended to the following: When~$s_j\geq s_j'$, $Y_j\vert S_j=s_j$ stochastically dominates~$Y_j\vert S_j=s_j'$ in the multivariate first-order stochastic dominance order (see, for instance, \cite{shaked2007}). The latter is equivalent to~$\mathbb P(Y_j\in U\vert S_j=s_j)\geq\mathbb P(Y_j\in U\vert S_j=s_j')$ for all upper sets~$U$ in~$\mathbb R^d$, where an upper set is defined by~$y\geq y'\,\&\,y'\in U\Rightarrow y\in U$. The strategy to derive bounds on potential outcomes and counterfactual parameters can be extended under the multivariate stochastic monotonicity assumption above.
This approach also allows us to derive sharp bounds on treatment effect parameters defined for multiple outcomes as discussed in \cite{mullahy2018}.

\subsection{Tighter bounds} 

One of the advantages of our approach is that additional identifying assumptions can easily be combined with Assumptions~\ref{ass:stability}-\ref{ass:monotonicity} to tighten the bounds on causal and counterfactual parameters. 

\textit{Factor structure:} 
We can obtain additional identifying power by reducing the degree of unobserved heterogeneity in potential outcomes. For example, one might assume that potential outcomes vary by university and field of study, but without systematic university-field interaction effects. More precisely, assume that potential outcomes can be written~$Y_j=V_{\upsilon(j)}+W_{\zeta(j)}+\epsilon_j$, where~$\upsilon(j)$ is the university hosting program~$j$, $\zeta(j)$ is the field of program~$j$, and the residual~$\epsilon_j$ has mean zero and is independent of~$(V_{\upsilon(j)}+W_{\zeta(j)},P,S)$. In case~$\mathcal I$ is completely unknown, this yields a tightening of the bounds on~$\mathbb E[Y_j\vert P=p,S_j=s_j]=\mathbb E[V_{\upsilon(j)}+W_{\zeta(j)}\vert P=p,S_j=s_j]$ obtained under Assumptions~\ref{ass:stability}-\ref{ass:monotonicity}.

\textit{Covariates and exclusion restrictions:} 
Covariates that affect potential outcomes in some programs but not others, such as the provision of an amenity specific to an education or labor market sector, provide additional identifying information in the form of intersection bounds. The latter can be straightforwardly combined with the intersection bounds from assumption~\ref{ass:monotonicity} to tighten the overall bounds on causal and counterfactual parameters.

\subsection{Inference}

Inference on our bounds for causal and counterfactual parameters relies on the now considerable literature on intersection bounds. The conditioning on rank-order lists~$P=p$ is a nonstandard aspect of our approach. Some cells in the support of the distribution of rank-order lists conditional on grade will have very few data points. Hence the need for smoothing in the space of permutations. We propose to do this with the Mallows kernel, which measures the discrepancy between two orderings~$p$ and~$p'$ of a set of~$n$ objects by~$K^\lambda(p,p'):=\exp\{-\lambda\, n_d(p,p')\}$, where~$\lambda\geq0$ is a tuning parameter and~$n_d(p,p')$ counts the number of discordant pairs. 
\cite{mallows2018} show that the Mallows kernel is positive definite, can be computed in~$O(J\log J)$ time and can be straightforwardly extended to partial orders, thereby providing a computationally attractive way of smoothing on~$p$.

%

\appendix
\numberwithin{equation}{section} 

\section{Proofs of results in the main text}
\label{app:proofs}

The proofs rely on the following properties of the extended order~$\succ_p$ defined in section~\ref{sec:bounds}.

\begin{lemma}\label{lem:order}
Maintain conventions (i)-(iii) and fix~$s$ and~$p\in\m{P}(s)$.
\begin{enumerate}[(a)]
\item $\succ_p$ is a strict partial order on~$\mJ^0$. For~$j,k\in p$, $j\succ_p k$ if and only if~$j\,p\,k$; for~$j\in p$ and~$k\in(\mathcal C(p)\setminus p)\cup\{0\}$, $j\succ_p k$; and no other pair of alternatives is ordered by~$\succ_p$.
\item $p(B(s))\succeq_p p(B^0(s))$; if~$p(B(s))\neq p(B^0(s))$, then~$p(B(s))\in p$ and~$p(B(s))\succ_p p(B^0(s))$.
\item Under assumption~\ref{ass:Roy}, $j\succeq_p k$ implies~$F_j(y\vert p,s)\leq F_k(y\vert p,s)$ for all~$y$; in particular, $F_{p(B(s))}(y\vert p,s)\leq F_{p(B^0(s))}(y\vert p,s)$ for all~$y$.
\end{enumerate}
\end{lemma}

\begin{proof}
\emph{Part (a).} The relation~$\succ_p$ is irreflexive: $j\succ_p j$ would require~$j\in p$ and either~$j\,p\,j$, which is excluded since the order of the list is strict, or~$j\notin p$. It is transitive: let~$j\succ_p k$ and~$k\succ_p l$. Then~$k\in p$ by definition of~$k\succ_p l$, so that~$j\succ_p k$ requires~$j\,p\,k$. Moreover, $l\in\mathcal C(p)\cup\{0\}$ and either~$k\,p\,l$, in which case~$j\,p\,l$ by transitivity of the order of the list, or~$l\notin p$. In both cases~$j\succ_p l$, since~$j\in p$ and~$l\in\mathcal C(p)\cup\{0\}$. Hence~$\succ_p$ is a strict partial order. If~$j,k\in p$, then~$k\in\mathcal C(p)$ by convention (i) and~$k\notin p$ fails, so that~$j\succ_p k$ if and only if~$j\,p\,k$. If~$j\in p$ and~$k\in(\mathcal C(p)\setminus p)\cup\{0\}$, then~$k\in\mathcal C(p)\cup\{0\}$ and~$k\notin p$, so that~$j\succ_p k$. Finally, $j\succ_p k$ requires~$j\in p$ and~$k\in\mathcal C(p)\cup\{0\}$, and the two previous cases exhaust such pairs.

\emph{Part (b).} Write~$a:=p(B(s))$ and~$b:=p(B^0(s))$. Since~$B^0(s)\subseteq B(s)$, $B^0(s)\cap p\subseteq B(s)\cap p$. If~$a=b$, there is nothing to prove. Otherwise, $B(s)\cap p\neq\emptyset$, since~$a=0$ would force~$B^0(s)\cap p=\emptyset$, hence~$b=0=a$, a contradiction; thus~$a\in p$ by convention (iii). If~$b\in p$, then~$b\in B(s)\cap p$ and~$b\neq a$, so that~$a\,p\,b$ by definition of~$a$, and~$a\succ_p b$ by part (a). If~$b=0$, then~$a\succ_p 0$ by part (a).

\emph{Part (c).} Let~$j\succeq_p k$. If~$j=k$, there is nothing to prove. If~$j\succ_p k$, then~$j\in p$, $k\in\mathcal C(p)\cup\{0\}$, and either~$j\,p\,k$ or~$k\notin p$, so that assumption~\ref{ass:Roy} yields~$Y_j(\omega)\geq Y_k(\omega)$ for every student~$\omega$ with~$P(\omega)=p$. Hence~$\{Y_j\leq y\}\subseteq\{Y_k\leq y\}$ on~$\{P=p\}$, and~$F_j(y\vert p,s)\leq F_k(y\vert p,s)$. The particular case follows from part (b).
\end{proof}

\begin{proof}[Proof of Theorem~\ref{thm:order}]
Fix~$s$, $p\in\m{P}(s)$, $y$ and~$j\in\mJ^0$. We first verify~(\ref{eq:pointid}). On the event~$\{P=p,S=s\}$, assumption~\ref{ass:stability} gives~$D=p(B(s))$, hence~$Y=Y_{p(B(s))}$, which is the first line of~(\ref{eq:pointid}). If~$p(B^0(s))\neq p(B(s))$, then~$p(B(s))\in B(s)\setminus B^0(s)$, i.e., $s=c_{p(B(s))}$, and the second line of~(\ref{eq:pointid}) is assumption~\ref{ass:continuity} applied at the cutoff~$c_k$ with~$k=p(B(s))$.

Next, for every~$k\in\{p(B(s)),p(B^0(s))\}$ such that~$j\succeq_p k$, Lemma~\ref{lem:order}(c) gives~$F_j(y\vert p,s)\leq F_k(y\vert p,s)$. Taking the minimum over all such~$k$ yields~$F_j(y\vert p,s)\leq U_j(y\vert p,s)$, where the convention~$\min\emptyset=1$ covers the case in which no such~$k$ exists. Symmetrically, for every~$k\in\{p(B(s)),p(B^0(s))\}$ such that~$k\succeq_p j$, Lemma~\ref{lem:order}(c) gives~$F_k(y\vert p,s)\leq F_j(y\vert p,s)$, and taking the maximum over all such~$k$ yields~$L_j(y\vert p,s)\leq F_j(y\vert p,s)$, with~$\max\emptyset=0$.

It remains to establish the closed forms. Write~$a:=p(B(s))$, $b:=p(B^0(s))$ and~$F_k:=F_k(y\vert p,s)$ for~$k\in\mJ^0$, so that~$a\succeq_p b$ by Lemma~\ref{lem:order}(b) and~$F_a\leq F_b$ by Lemma~\ref{lem:order}(c). Consider first~$U_j(y\vert p,s)$. If~$j\succeq_p a$, then~$j\succeq_p b$ by transitivity, so that the minimum in~(\ref{eq:roy-bounds1}) runs over~$\{F_a,F_b\}$ and equals~$F_a$. If~$a\succ_p j\succeq_p b$, then~$j\succeq_p a$ fails by asymmetry, so that the minimum runs over~$\{F_b\}$. Otherwise, we claim that~$j\succeq_p b$ fails; then~$j\succeq_p a$ fails as well, since it would imply~$j\succeq_p b$, and the minimum runs over the empty set and equals~$1$. To prove the claim, suppose that~$j\succeq_p b$ and~$j\not\succeq_p a$. Then~$a\neq b$, hence~$a\in p$ and~$a\succ_p b$ by Lemma~\ref{lem:order}(b). If~$j=b$, then~$a\succ_p j$. If~$j\succ_p b$, then~$j\in p$ and~$j\neq a$, so that Lemma~\ref{lem:order}(a) gives either~$j\,p\,a$, i.e., $j\succ_p a$, which is excluded, or~$a\,p\,j$, i.e., $a\succ_p j$. In both cases, $a\succ_p j\succeq_p b$, which is the second case. Consider now~$L_j(y\vert p,s)$. If~$b\succeq_p j$, then~$a\succeq_p j$ by transitivity, so that the maximum in~(\ref{eq:roy-bounds2}) runs over~$\{F_a,F_b\}$ and equals~$F_b$. If~$a\succeq_p j$ and~$b\not\succeq_p j$, the maximum runs over~$\{F_a\}$. Otherwise, it runs over the empty set and equals~$0$.
\end{proof}

\begin{proof}[Proof of Theorem~\ref{thm:bounds}]
Fix~$j\in\mJ$, $y$, $s$ and~$p\in\m{P}(s)$, and write~$\pi_{s'}(p'):=\mmp[P=p'\vert S=s']$, which is positive for~$p'\in\m{P}(s')$. Since~$S$ is scalar and common to all programs, assumption~\ref{ass:monotonicity} states that~$s'\mapsto\mmp[Y_j\leq y\vert S=s']$ is nonincreasing, which is~(\ref{eq:mon}). Since the set of rank-order lists is finite, the law of total probability gives, for every~$s'$,
\begin{eqnarray}\label{eq:lotp}
\mmp[Y_j\leq y\vert S=s'] & = & \sum_{p'\in\m{P}(s')}\pi_{s'}(p')F_j(y\vert p',s').
\end{eqnarray}
Applying Theorem~\ref{thm:order} at each~$(p',s')$ in~(\ref{eq:lotp}) and combining with~(\ref{eq:mon}) yields
\begin{equation}\label{eq:envelope}
\begin{aligned}
\sup_{s'\geq s}\sum_{p'\in\m{P}(s')}\pi_{s'}(p')L_j(y\vert p',s') & \;\leq\; \mmp[Y_j\leq y\vert S=s]\\
& \;\leq\; \inf_{s'\leq s}\sum_{p'\in\m{P}(s')}\pi_{s'}(p')U_j(y\vert p',s').
\end{aligned}
\end{equation}
Isolating the term~$p'=p$ in~(\ref{eq:lotp}) at~$s'=s$ gives
\begin{eqnarray*}
\pi_s(p)F_j(y\vert p,s) & = & \mmp[Y_j\leq y\vert S=s]-\sum_{p'\in\m{P}(s)\setminus\{p\}}\pi_s(p')F_j(y\vert p',s).
\end{eqnarray*}
Bounding the first term on the right hand side by~(\ref{eq:envelope}), each term of the sum by Theorem~\ref{thm:order} at~$(p',s)$, and dividing by~$\pi_s(p)>0$, yields~$\tilde L_j(y\vert p,s)\leq F_j(y\vert p,s)\leq\tilde U_j(y\vert p,s)$, with~$\tilde L_j$ and~$\tilde U_j$ as in~(\ref{eq:mon-bounds}), since~$\pi_{s'}(p')/\pi_s(p)=w^{p'\vert s'}_{p\vert s}$. Together with Theorem~\ref{thm:order}, this is~(\ref{eq:final}).
\end{proof}

\begin{proof}[Proof of Theorem~\ref{thm:cf}]
Fix~$y$. The identified set for~$\theta(y)$ is the set of values of~(\ref{eq:theta}) across models~$(Y_0,\ldots,Y_J,P,S)$ that satisfy Assumptions~\ref{ass:stability}-\ref{ass:Roy} and generate the distribution of~$(P,S,Y)$. Call a family~$\{F_j(\cdot\vert p,s): j\in\mJ^0, s, p\in\m{P}(s)\}$ of CDFs, measurable in~$s$, \emph{admissible} if (F1) $F_{p(B(s))}(\cdot\vert p,s)$ and~$F_{p(B^0(s))}(\cdot\vert p,s)$ are given by~(\ref{eq:pointid}), and (F2) $j\succeq_p k$ implies~$F_j(\cdot\vert p,s)\leq F_k(\cdot\vert p,s)$. Let~$\m{F}^R$ be the set of admissible families, and note that~$\theta(y)$ in~(\ref{eq:theta}) is an affine function of the family of conditional CDFs.

\emph{Step 1: the identified set is the image of~$\m{F}^R$ by~$\theta(y)$, and it is an interval.} Let~$(Y_0,\ldots,Y_J,P,S)$ satisfy Assumptions~\ref{ass:stability}-\ref{ass:Roy}. Its family of conditional CDFs satisfies (F1) by Assumptions~\ref{ass:stability}-\ref{ass:continuity}, as shown in the proof of Theorem~\ref{thm:order}, and (F2) by Lemma~\ref{lem:order}(c). Conversely, let~$\{F_j(\cdot\vert p,s)\}$ be admissible and construct a model as follows: $(P,S)$ has the observed joint distribution; conditionally on~$(P,S)=(p,s)$, draw~$V$ uniform on~$(0,1)$ and set~$Y_j:=F_j^{-1}(V\vert p,s)$ for~$j\in\mJ^0$, where~$F^{-1}(v):=\inf\{u: F(u)\geq v\}$; set~$D:=P(B(S))$ and~$Y:=Y_D$. In this model, $Y_j$ has conditional CDF~$F_j(\cdot\vert p,s)$ given~$(P,S)=(p,s)$, assumption~\ref{ass:stability} holds by construction, and~$Y$ has conditional CDF~$F_{p(B(s))}(\cdot\vert p,s)$ given~$(P,S)=(p,s)$, which is the observed one by (F1), so that the model generates the distribution of~$(P,S,Y)$. If~$j\succ_p k$, then~$F_j(\cdot\vert p,s)\leq F_k(\cdot\vert p,s)$ by (F2), hence~$F_j^{-1}(v\vert p,s)\geq F_k^{-1}(v\vert p,s)$ for all~$v$ and~$Y_j\geq Y_k$ on~$\{P=p\}$: this is assumption~\ref{ass:Roy}. Let~$c_k$ be a cutoff, $p$ a list and~$j:=p(B^0(c_k))\neq p(B(c_k))$. Since there are finitely many cutoffs, $B(s')=B^0(c_k)$ for all~$s'<c_k$ close enough to~$c_k$, hence~$p(B(s'))=j$ and~$\mmp[Y_j\leq u\vert P=p,S=s']=\mmp[Y\leq u\vert P=p,S=s']$ for such~$s'$ and all~$u$; by (F1), the limit of the latter as~$s'\uparrow c_k$ is~$F_j(u\vert p,c_k)=\mmp[Y_j\leq u\vert P=p,S=c_k]$, which is assumption~\ref{ass:continuity}. Hence the identified set is the image of~$\m{F}^R$ by~$\theta(y)$. The set~$\m{F}^R$ is convex, since (F1) and (F2) are preserved by convex combinations, and~$\theta(y)$ is affine in the family, so that this image is an interval.

\emph{Step 2: validity.} Let the family belong to~$\m{F}^R$ and fix~$(p,s)$. By (F2), $F_j(y\vert p,s)\leq F_k(y\vert p,s)$ for every~$k\in\{p(B(s)),p(B^0(s))\}$ with~$j\succeq_p k$, and these~$F_k(y\vert p,s)$ are given by (F1), so that~$F_j(y\vert p,s)\leq U_j(y\vert p,s)$; symmetrically~$F_j(y\vert p,s)\geq L_j(y\vert p,s)$. Substituting in~(\ref{eq:theta-movers}) gives~$\underline\theta^R(y)\leq\theta(y)\leq\overline\theta^R(y)$, and by Step 1 it remains to show that both endpoints are attained.

\emph{Step 3: the endpoints are attained.} Write~$a:=p(B(s))$, $b:=p(B^0(s))$ and~$\m{A}:=\{j\in\mJ^0: j\succeq_p b\}$, which contains every~$j\succeq_p a$ since~$a\succeq_p b$ by Lemma~\ref{lem:order}(b), and consider the family
\[
F_j(\cdot\vert p,s):=U_j(\cdot\vert p,s)\;\mbox{ for }j\in\m{A},\qquad F_j(\cdot\vert p,s):=\max\{F_b(\cdot\vert p,s),\mmi\{\cdot\geq y\}\}\;\mbox{ for }j\notin\m{A},
\]
where~$U_j(\cdot\vert p,s)\in\{F_a(\cdot\vert p,s),F_b(\cdot\vert p,s)\}$ for~$j\in\m{A}$ by the closed form in Theorem~\ref{thm:order}. All members are CDFs, and (F1) holds since~$a,b\in\m{A}$, $U_a=F_a$ and~$U_b=F_b$. For (F2), let~$j\succ_p k$. If~$k\in\m{A}$, then~$j\in\m{A}$ by transitivity, and~$U_j\leq U_k$ because the minimum in~(\ref{eq:roy-bounds1}) runs over a larger set for~$j$ than for~$k$. If~$j\in\m{A}$ and~$k\notin\m{A}$, then~$U_j\leq F_b\leq F_k$, since~$F_a\leq F_b$ by Lemma~\ref{lem:order}(c). If~$j,k\notin\m{A}$, then~$F_j=F_k$. Finally, $F_j(y\vert p,s)=U_j(y\vert p,s)$ for all~$j$, since~$U_j(y\vert p,s)=1$ for~$j\notin\m{A}$. The family is measurable in~$s$, belongs to~$\m{F}^R$, and yields~$\theta(y)=\overline\theta^R(y)$. For the lower endpoint, let~$\m{A}':=\{j\in\mJ^0: a\succeq_p j\}$, which contains every~$j$ with~$b\succeq_p j$, and set~$F_j(\cdot\vert p,s):=L_j(\cdot\vert p,s)\in\{F_a(\cdot\vert p,s),F_b(\cdot\vert p,s)\}$ for~$j\in\m{A}'$ and~$F_j(\cdot\vert p,s):=\min\{F_a(\cdot\vert p,s),\mmi\{\cdot\geq y+1\}\}$ for~$j\notin\m{A}'$. If~$j\succ_p k$ and~$j\in\m{A}'$, then~$k\in\m{A}'$ and~$L_j\leq L_k$, because the maximum in~(\ref{eq:roy-bounds2}) runs over a smaller set for~$j$ than for~$k$; if~$j\notin\m{A}'$ and~$k\in\m{A}'$, then~$F_j\leq F_a\leq L_k$; if~$j,k\notin\m{A}'$, then~$F_j=F_k$. Hence the family belongs to~$\m{F}^R$, and~$F_j(y\vert p,s)=L_j(y\vert p,s)$ for all~$j$, since~$L_j(y\vert p,s)=0$ for~$j\notin\m{A}'$, so that~$\theta(y)=\underline\theta^R(y)$.
\end{proof}

\section{Additional results}
\begin{proposition}\label{prop:iteration}
Suppose Assumptions~\ref{ass:stability}-\ref{ass:monotonicity} hold. Fix~$j\in\mJ$ and~$y$, and let~$L^{(1)}_j(y\vert p,s)$ and~$U^{(1)}_j(y\vert p,s)$ denote the lower and upper bounds in~(\ref{eq:final}). Let~$\tilde L^{(1)}_j$ and~$\tilde U^{(1)}_j$ be defined as~$\tilde L_j$ and~$\tilde U_j$ with~$(L_j,U_j)$ replaced by~$(L^{(1)}_j,U^{(1)}_j)$. Then~$\max\{L^{(1)}_j,\tilde L^{(1)}_j\}=L^{(1)}_j$ and~$\min\{U^{(1)}_j,\tilde U^{(1)}_j\}=U^{(1)}_j$ for all~$s$ and~$p\in\m{P}(s)$.
\end{proposition}

\begin{proof}
We prove the statement for the upper bound; the proof for the lower bound is symmetric. Write~$\pi_s(p):=\mmp[P=p\vert S=s]$, $G(s):=\mmp[Y_j\leq y\vert S=s]$, $G^U(s):=\sum_{p\in\m P(s)}\pi_s(p)U_j(y\vert p,s)$, $G^L(s):=\sum_{p\in\m P(s)}\pi_s(p)L_j(y\vert p,s)$, $m(s):=\inf_{s'\leq s}G^U(s')$ and~$M(s):=\sup_{s'\geq s}G^L(s')$, so that
\begin{eqnarray*}
\pi_s(p)\tilde U_j(y\vert p,s) & = & m(s)-\sum_{p'\in\m P(s)\setminus\{p\}}\pi_s(p')L_j(y\vert p',s),\\
\pi_s(p)\tilde L_j(y\vert p,s) & = & M(s)-\sum_{p'\in\m P(s)\setminus\{p\}}\pi_s(p')U_j(y\vert p',s).
\end{eqnarray*}
Under Assumptions~\ref{ass:stability}-\ref{ass:monotonicity}, $G^L(s)\leq G(s)\leq G^U(s)$ and~$M(s)\leq G(s)\leq m(s)$ for all~$s$, and~$m$ and~$M$ are nonincreasing.

\emph{Step 1: the envelope is unchanged.} Let~$G^{U,(1)}(s):=\sum_{p\in\m P(s)}\pi_s(p)U^{(1)}_j(y\vert p,s)$ and~$A:=\{p\in\m P(s):\tilde U_j(y\vert p,s)<U_j(y\vert p,s)\}$. If~$A=\emptyset$, then~$G^{U,(1)}(s)=G^U(s)\geq m(s)$. Otherwise, pick~$p^*\in A$. For~$p\in A$, $\pi_s(p)\tilde U_j(y\vert p,s)=m(s)-G^L(s)+\pi_s(p)L_j(y\vert p,s)\geq \pi_s(p)L_j(y\vert p,s)$, since~$G^L(s)\leq m(s)$. Hence
\begin{eqnarray*}
G^{U,(1)}(s) & \geq & \sum_{p\notin A}\pi_s(p)U_j(y\vert p,s)+\left(m(s)-\sum_{p'\neq p^*}\pi_s(p')L_j(y\vert p',s)\right)+\sum_{p\in A\setminus\{p^*\}}\pi_s(p)L_j(y\vert p,s)\\
& = & m(s)+\sum_{p\notin A}\pi_s(p)\left[U_j(y\vert p,s)-L_j(y\vert p,s)\right]\;\geq\; m(s).
\end{eqnarray*}
Since~$G^{U,(1)}\leq G^U$ pointwise and~$m$ is nonincreasing, $\inf_{s'\leq s}G^{U,(1)}(s')=m(s)$ for all~$s$. Symmetrically, $\sup_{s'\geq s}G^{L,(1)}(s')=M(s)$, where~$G^{L,(1)}(s):=\sum_{p\in\m P(s)}\pi_s(p)L^{(1)}_j(y\vert p,s)$.

\emph{Step 2: the residual term does not help.} By step 1,
\begin{eqnarray*}
\pi_s(p)\tilde U^{(1)}_j(y\vert p,s) & = & m(s)-\sum_{p'\neq p}\pi_s(p')L^{(1)}_j(y\vert p',s).
\end{eqnarray*}
Let~$A':=\{p'\in\m P(s)\setminus\{p\}: \tilde L_j(y\vert p',s)>L_j(y\vert p',s)\}$. If~$A'=\emptyset$, then~$\tilde U^{(1)}_j(y\vert p,s)=\tilde U_j(y\vert p,s)$. Otherwise, pick~$p_1\in A'$. For~$p'\in A'$, since~$M(s)\leq G^U(s)$,
\begin{eqnarray*}
\pi_s(p')L^{(1)}_j(y\vert p',s) & = & M(s)-\sum_{p''\neq p'}\pi_s(p'')U_j(y\vert p'',s)\;\leq\; \pi_s(p')U_j(y\vert p',s).
\end{eqnarray*}
Hence
\begin{eqnarray*}
\sum_{p'\neq p}\pi_s(p')L^{(1)}_j(y\vert p',s) & \leq & \left(M(s)-\sum_{p''\neq p_1}\pi_s(p'')U_j(y\vert p'',s)\right)+\sum_{p'\in A'\setminus\{p_1\}}\pi_s(p')U_j(y\vert p',s)\\
& & \;+\;\sum_{p'\notin A'\cup\{p\}}\pi_s(p')L_j(y\vert p',s)\\
& = & M(s)-\pi_s(p)U_j(y\vert p,s)-\sum_{p'\notin A'\cup\{p\}}\pi_s(p')\left[U_j(y\vert p',s)-L_j(y\vert p',s)\right]\\
& \leq & M(s)-\pi_s(p)U_j(y\vert p,s),
\end{eqnarray*}
so that~$\pi_s(p)\tilde U^{(1)}_j(y\vert p,s)\geq m(s)-M(s)+\pi_s(p)U_j(y\vert p,s)\geq\pi_s(p)U_j(y\vert p,s)$. In both cases, $\tilde U^{(1)}_j(y\vert p,s)\geq\min\{U_j(y\vert p,s),\tilde U_j(y\vert p,s)\}=U^{(1)}_j(y\vert p,s)$, which proves the claim for the upper bound.
\end{proof}

\begingroup
\makeatletter
\def,{%
  \ifmmode%
    \old@comma\discretionary{}{}{}%
  \else%
    \unskip\old@comma%
  \fi%
}
\makeatother
\bibliographystyle{ecta}
\bibliography{Roy+RD}
\endgroup

\end{document}